\documentclass[journal]{IEEEtran}

\usepackage[T1]{fontenc}
\usepackage{flushend}

\ifCLASSINFOpdf
\else
\fi
\usepackage{amsmath}
\usepackage{graphicx}
\usepackage{subfigure}
\usepackage{amsmath}
\usepackage{amssymb}
\usepackage{balance}
\usepackage{paralist}
\begin{document}
%
\title{Catalyzing Cloud-Fog Interoperation in 5G Wireless Networks: An SDN Approach}
%
\author{Peng~Yang,~Ning~Zhang,~Yuanguo~Bi,~Li~Yu,~and~Xuemin~(Sherman)~Shen%
\thanks{Peng Yang and Li Yu (corresponding author) are with Huazhong University of Science and Technology. Ning Zhang and Xuemin (Sherman) Shen are with University of Waterloo. Yuanguo Bi is with Northeastern University, China.}}%

\maketitle
 
\begin{abstract}
The piling up storage and compute stacks in cloud data center are expected to accommodate the majority of internet traffic in the future. However, as the number of mobile devices significantly increases, getting massive data into and out of the cloud wirelessly inflicts high pressure on the bandwidth, and meanwhile induces unpredictable latency. Fog computing, which advocates extending clouds to network edge, guarantees low latency and location-aware service provisioning. In this article, we consider fog computing as an ideal complement rather than a substitute of cloud computing, and we propose a software defined networking (SDN) enabled framework for cloud-fog interoperation, aiming at improving quality of experience and optimizing network resource usage. Two case studies are provided to illuminate the feasibility and advantage of the proposed framework. At last, potential research issues are presented for further investigation.
\end{abstract}


\IEEEpeerreviewmaketitle

\section{Introduction}
Recently, service providers (SPs) are enthusiastically exploring cloud computing for service provisioning. Typically, cloud vendors provide large scale storage and compute resource pooling, enabling SPs to customize their level of service on a pay-as-you-go basis. SPs are then able to make faster innovation with reduced capital expenditure and operational expenditure. Consequently, more and more services are being delivered from the cloud. However, this trend will be shifted when mobile device becomes the dominant user equipment. Cisco forecasts that, by 2020, there will be 1.5 mobile devices per capita globally \cite{cisco VNI}. As mobile devices popularity surges, the superiority of cloud computing will be adversely affected since getting massive data into and out of the cloud wirelessly requires substantial spectrum resources and incurs unexpected latency, which lead to degraded quality of experience (QoE).

To provide users with better experience in the coming 5G era, massive MIMO and adaptive access will be exploited for ubiquitous connection and higher data rate. Meanwhile, cloud-assisted platform is also designed to enhance the operation and management of heterogeneous 5G networks \cite{cloud assisted hetnet}. Among others, \emph{Fog computing} reveals an alternative approach to ease the wireless networking tensions \cite{tom}. Instead of migrating data between cloud and mobile users, fog computing advocates extending data centers from network center to network edge, which tremendously reduces the data volume traversed through core network. Meanwhile, the round-trip latency can be shortened considerably. Moreover, by implementing fog computing in the vicinity of mobile users, geo-location featured applications (e.g., vehicular network, e-healthcare system and smart home system) can be better supported since data generated in those scenarios will most likely to be consumed locally. Yet, fog computing is an ideal complement rather than a substitute of cloud computing. As mobile service type diversifies, user requirements can hardly be met by monotone computing paradigm. Certain services can be better delivered from cloud while others from fog. In this way, fog and cloud are interdependent and should be investigated together. However, harnessing them unitedly is challenging due to the following critical questions: What kind of tasks should be processed at fog? How should fog interoperate with cloud to guarantee QoE and, meanwhile, maximize the usage of fog resources?

Software defined networking (SDN), characterized by the decoupled control plane and data plane, provides fine-grained network control. Based on real-time global information, the SDN controller is able to make informed management decisions \cite{sdnsurvey}. Because of this, SDN is witnessing growing popularity in complex network management. Liu \emph{et al.} take the advantage of SDN to conduct fine-grained measurement and control in device-to-device based multi-tier LTE-A networks, aiming at providing better QoE \cite{d2d}. In this article, we propose an SDN enabled architecture to catalyze the cloud-fog interoperation. Through the logical controller and programmable interfaces elaborated in our architecture, the full potential of local pooling resources can be unleashed. Meanwhile, user requests can be dynamically steered among fogs and cloud so that they can be processed with better QoE. Moreover, the proposed architecture will also benefit network operators in terms of agile management and cost-effectiveness. 

The remainder of the article is organized as follows. We introduce fog computing and its applications in the next section, and clarify the importance of cloud-fog interoperation and its challenges in the following section. Then, we propose an SDN enabled architecture for cloud-fog interoperation. Two case studies are conducted later to highlight the feasibility and advantage of the proposed architecture. We present several potential research issues, and finally the conclude the article in the last section.

\section{Why Fog Computing in 5G System?}\label{whyfog}
Cloud computing provides outsourced infrastructures, platforms and softwares. Being able to deliver such full-scale services on demand, clouds are powerful and cost-effective. However, there are inherent limitations that mobile applications will suffer from the cloud paradigm in 5G era.

\subsection{Cloud Limitations}
\subsubsection{The Responsiveness}
The coming 5G system is envisioned to provide $1$ millisecond (ms) round-trip time. This vision, however, might be hard to be carried out since that, packets of mobile users have to traverse radio access network (RAN), core network (CN) and internet before reaching the cloud server. Current LTE networks achieves $10$ ms round-trip latency, of which $5$ ms in RAN and CN, and $5$ ms in the internet if the server is in the same country as the user \cite{GSMA Latency}. 
Even the time spent in RAN and CN is managed to be within $1$ ms in 5G system, the time spent in the internet turned out to be the dominant barrier. Moreover, bandwidth limitation and network uncertainty on the way to clouds will further make the latency performance unpredictable.

\subsubsection{The Backhaul Bottleneck}
Through densification, as well as advanced physical layer techniques, 5G system is supposed to provide seamless coverage and gigabit data rate. However, even if those measures are adopted to make users stay connected, high data rate might still be hard to achieve due to the backhaul bottleneck. Since user requested contents have to be delivered from cloud servers through backhaul, the capacity of backhaul should be at least comparable to that of RAN, which is challenging.

\subsubsection{Location-Aware Applications}
The strong momentum of location based services is another driving force towards fog computing. In those applications, services are often requested by geographically adjacent users, and data exchange mostly happens locally. Unnecessarily direct those traffic to the cloud will not only impair the responsiveness, but also aggravate the backhaul tension. If those traffic can be handled around the application contexts, users can be served with better experience and cost effectiveness.

\subsection{Fog Computing}
Given those cloud limitations, researchers are extensively investigating providing services near mobile users. The measure they have taken is twofold: delivering contents near users and offloading users' computation tasks to nearby powerful agents. Ao \emph{et al.} propose to cache contents at small base stations, and they present cooperative algorithms that achieve unprecedented content delivery speed at reduced backhaul cost \cite{mobihoc}. Chen \emph{et al.} investigate the distributed computation offloading game for mobile edge computing considering latency and energy consumption \cite{XuChen}. These works and references therein exploit either storage or compute resources around users to improve QoE. However, emerging applications like virtual/augmented reality generate large volume of data that need to be processed instantly. It poses challenges to mobile devices in terms of storage, communication and computation.

\begin{figure}[]
  \centering
  \includegraphics[width=0.48\textwidth]{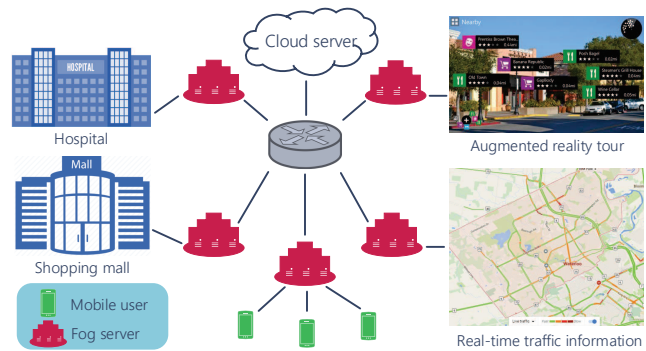}
  \caption{The prospective fog applications.}\label{fogs}
\end{figure}

Fog computing can help users to well meet those requirements. By pooling resources from near-user devices and infrastructures, fog computing carries out substantial storage and computation capacity with reduced communication cost. In this way, heavy applications can be delivered with shortened latency, energy efficiency and location awareness \cite{vfog, fogJSAC}. Fascinating fog services can be expected, for example, caching popular contents at hotspots can largely improve the responsiveness. Meanwhile, fog can also be harnessed for location-specific purpose. Fig. \ref{fogs} depicts several fog contexts: fog server in a shopping mall can provide commodity information and navigation to nearby consumers; community-wide fog can be utilized to collect household utility usage for demand analysis of utility companies; city-wide fog can be leveraged for gathering, analyzing and disseminating real-time road traffic information. The granularity of fogs can vary from body area to city wide.

\section{Cloud-Fog Interoperation and Challenges}\label{interoperation}
\subsection{The Interoperation}
In fact, cloud can easily host various fog applications, even with more than enough resources, but the unpredictable latency and bandwidth wastage is unacceptable. Though fog outperforms cloud in terms of responsiveness and location awareness, its lightweight capability will be inadequate when handling heavy tasks. Therefore, fog computing is complementary, rather than substitutional, to cloud computing. Computationally intensive tasks, long-term large-size file storing and sharing, delay-tolerant service are more appropriate to be handled by the cloud. While the massively deployed fogs can well carry applications like lightweight computing and short lived location-specific content sharing.

More importantly, we are seeing huge potential of cloud-fog cooperation. Take augmented reality tour as an example. While traveling, visitors can enjoy the sights with their smart phones or glasses. Extra information on landmarks or restaurants is displayed on the screen immediately. In this application, images are continuously generated and need to be processed instantly. By cloud-fog cooperation, image features can be firstly extracted at local fog, thereafter, the feature information is delivered to the cloud for image matching. In this way, bulk images are processed locally, which avoids bandwidth wastage and unexpected latency. Moreover, as cloud server stores massive data sets, retrieving image in the cloud will return better results.

\subsection{The Challenges}
Since requests and contents will migrate frequently between fogs and clouds, high-level cloud-fog interoperation should be enabled for coordinated service provisioning. However, empowering the interoperation is of great challenge due to the following requirements.

It needs a \textbf{local coordinator}. As service requirements are dynamic and time-varying, certain application may need support from either cloud or fog, or both of them. Thus, we need a coordinator to preprocess the requests. If the request can be cooperatively processed locally, the coordinator will need to decompose and assign tasks to available resources. If the request is more suitable to be processed in the cloud, the coordinator can act as local hub to direct it to the cloud.

It needs \textbf{global knowledge}. Cloud and fog might be operated by several vendors, and they have to accommodate the SPs coordinately. Therefore, gathering high-level and real-time information regarding these facilities is a key enabler to cloud-fog interoperation. Once the coordinator is well informed, it can make optimal decisions. Furthermore, a network state information renewal mechanism should be designed, as well as how to manage and make the best use of collected information.

It needs \textbf{open programmable interfaces}. Typical applications will extensively exchange information among mobile users, fog/cloud servers and service providers. In this way, each entity needs to make certain functions available to others through open interfaces, so that high-level policies from controller can be dynamically enforced to different entities.

\section{SDN Enabled Cloud-Fog Computing Architecture}\label{archit}
In legacy networks, control and data plane are tightly coupled, protocols running in switches and routers are immutable once they are installed. Hence the network is highly ossified and the operator can hardly innovate upon it. SDN, a new network paradigm that enables control-data separated operation, gives the controller a global view of the network via programmable control plane. Consequently, it contributes to dynamic network deployment, agile network management, faster application innovation and efficient resource utilization \cite{sdnsurvey,d2d}. Cultivating SDN to catalyze the cloud-fog interoperation can well meet the requirements mentioned in the previous section. Therefore, we propose an SDN enabled architecture for cloud-fog interoperation.

\subsection{Architectural Overview}

\begin{figure}[]
  \centering
  \includegraphics[width=0.49\textwidth]{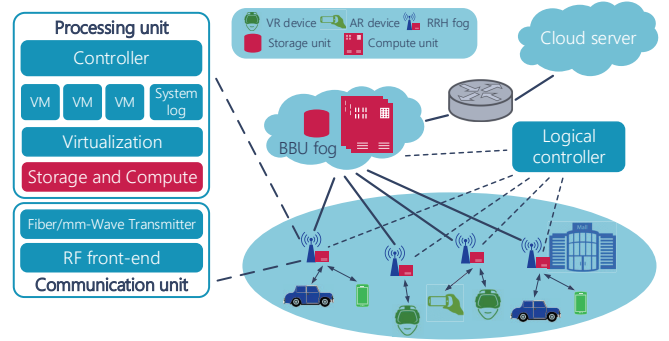}
  \caption{The SDN enabled cloud-fog interoperation architecture.}\label{architecture}
\end{figure}
\subsubsection{Components}
As illustrated in Fig. \ref{architecture}, the proposed architecture incorporates mobile users, fog servers and the cloud server, where storage and compute resources beyond access network are referred to as cloud. Fogs are deployed on top of cloud-RAN (C-RAN). C-RAN is an ideal carrier for fog deployment, it divides the function of conventional base stations into two parts: remote radio head (RRH) for radio signal transceiving and baseband unit (BBU) for high-speed baseband processing \cite{cran tutorial}. By stacking storage and compute resources on BBU and RRHs, C-RAN can well carry the facilities for fog computing.

RRH fogs are widely scattered and attached to the centralized BBU fog (20-40 km away \cite{cran tutorial}) via fiber link or millimeter-wave. In addition to the communication unit, each RRH fog is equipped with a processing unit. The storage and compute resources are virtualized as isolated virtual machines (VMs) \cite{virtualization}, which is managed by a local fog controller. The functionalities of RRH fog lies in the following two folds: 1) it performs baseband processing after radio signals are received and 2) it provides local storage and compute resources that can be dynamically scheduled by the controller at BBU fog.

BBU fog is stacked with more processing units, making it the most powerful fog within RAN. In addition to the functionalities of RRH fog, the BBU controller also acts as a master controller (MC) that coordinates all the controllers in RRHs. Note that this configuration is compatible with the legacy C-RAN, and thus can be incrementally deployed.

\subsubsection{Interoperation}
With the coordination of MC, local network information is continuously shared among controllers, leading to a logic controller with group intelligence. The controllers evaluate user requests and make corresponding response through either fog-fog interoperation or cloud-fog interoperation.

Fog-fog interoperation is initiated when user demands is beyond the capability of an individual fog. Upon receiving such requests, the logical controller starts a crowdsourcing process \cite{crowdsourcing}. Based on the available resource on each fog server, the MC decomposes the task and distributes them accordingly. Once the decomposed tasks are all completed, the MC recomposes the results and delivers them to mobile users.

Cloud-fog interoperation is activated when user demands can not be met only by fog computing. In this regard, the MC abstracts the user requirement and evaluates whether fog preprocessing will accelerate the service provisioning, recall the example of augmented reality tour mentioned before. Then the MC either starts a crowdsourcing process through fog-fog interoperation or sends user requests directly to the cloud.

With interoperation, more user requests can be handled locally, and meanwhile fog resource efficiency is pushed to the maximum. Taking fog caching as an example, frequently requested files can be pushed to fogs from the cloud. Thus, mobile users can fetch contents from neighbouring servers through fog-fog interoperation. As fog storage is limited compared to the cloud, only selected contents can be cached and dated files should be evicted continuously. Thus, to obtain a higher local request hit ratio, cloud and fog have to interact frequently to maintain the optimality of fog content entry.

\subsection{Logical controller}

\begin{figure}[]
\centering
\includegraphics[width=0.4\textwidth]{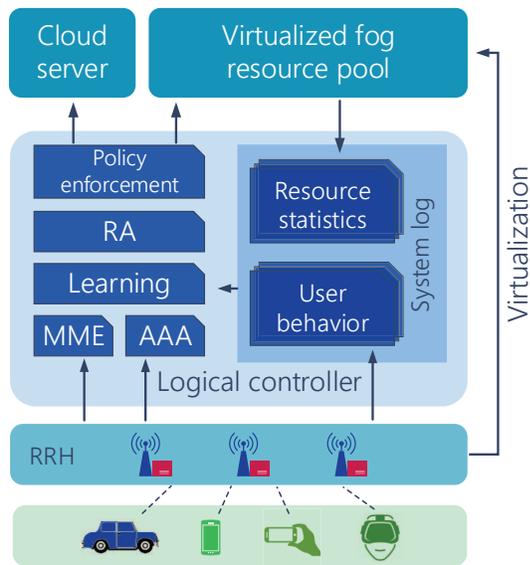}
\caption{The elements of logical controller for service provisioning.}\label{generalserviceprocess}
\end{figure}

Fig. \ref{generalserviceprocess} depicts the framework of logical controller geared service provisioning. The logical controller is responsible for bridging the mobile users and fog service providers. On one hand, it processes the requests from mobile users, and anonymously records them in the system log. On the other hand, it periodically updates the events and VM operating information of each fog in the system log, such as pending tasks, amount of users served and remaining resources. The controller thus has a global knowledge of the network status and can always make informed scheduling decisions based on what it has learned from the system log.

As the local network operator, the logic controller also manages the access network including authentication, authorization and accounting (AAA), mobility management entity (MME), and resource allocation (RA). The system log is essential for controller's decision making as it provides information of user requests and available resources. More importantly, those historical statistics can be exploited for learning and predicting the future network behaviors so as to take proactive actions. The prediction results are expected to be more accurate as the system log files expand.

\subsection{The Programmable Interfaces}
Generally, delivering services involves three other entities except the controller: mobile users, fog/cloud servers and SPs, while all of them are intertwined. A practical case is that, music streaming company Spotify uses Amazon's simple storage service to store its music files while keeping its core computing infrastructure (or its backend) in Google's cloud platform to process and record user requests\footnote{http://www.wsj.com/articles/google-cloud-lures-amazon-web-services-customer-spotify-1456270951}. Since information is extensively exchanged among those entities, standardized application programming interfaces (API) are pivotal. In computer networks, OpenFlow is the dominated software interface between SDN controller and the underlying switches. Extension of OpenFlow can be made to support the communication between the logical controller and fog facilities \cite{openroad}. Generally, APIs can be classified into two categories:
\begin{itemize}
  \item Functional APIs. Those APIs make functionalities at each entity available to others for service delivering, such as data migration, VM configuration and scheduling;
  \item Management APIs. Those APIs are used for AAA, mobility management, service billing and so on.
\end{itemize}

Once those APIs are elaborated and standardized, user's requirements can be cooperatively fulfilled by heterogeneous SPs with minimum impact on the performance.

\subsection{Advantages and Visions}
The innate SDN features of the proposed architecture are key enablers towards cloud-fog interoperation. Specifically, they bring the following advancements:
\begin{itemize}
  \item Real-time knowledge. Real-time network status information is essential to controller's decision making. The dedicated control plane makes the update of local information reliable and flexible.
  \item Centralized operation. The logical controller schedules and optimizes the utilization of all the resources within RAN. Eventually, the norm of wireless networks will be locally centralized and globally distributed.
  \item Fine-grained control. The dedicated control channel exposes traditionally inaccessible functional APIs to the controller. In turn, physical devices translate high-level policies into low-level configuration instructions, whereby fine-grained control is achieved.
\end{itemize}

Consequently, the proposed architecture has great potential to enhance the network performance in the following aspects.

\textbf{Improved QoE}. Being able to deliver services in the vicinity of mobile users, the SDN enabled cloud-fog interoperation architecture will largely reduce latency and jitter compared to the standalone cloud solution. The architecture will offer users with dramatically improved responsiveness and make the $1$ ms latency achievable in the coming 5G era.

As human-computer interface develops, users are now expecting to get information and entertainment in highly interactive ways. Exciting applications include deeply immersive augmented reality and virtual reality. In those contexts, users' devices interact constantly with application servers. For instances, as user's eyes roll, the virtual reality devices should show the exact sight that the user is looking at. Though computationally demanding, application servers have to make continuous prompt response, otherwise user experience will be easily deteriorated since humans are extremely sensitive to visual stutter. As shown later in the case study, the proposed architecture can incorporate those applications with improved QoE.

\textbf{Resource pooling economics}. In spite of the lightweight resources deployed in the fog, centralized operation and virtualization drive the multiplexing efficiency of resources to the maximum. In addition, since the traffic traversing backhaul is reduced, the expense on the backhaul facilities will also scale down. Moreover, wireless carriers have kept investing in deploying, operating and upgrading the network for decades. However, as value-added services are mostly delivered by internet SPs, wireless carriers have long been suffering from shrinking profits and, meanwhile, coping with the endless insatiable demands for higher data rate. This situation will be turned around if value-added services can be delivered within RAN, since carriers will have decided advantage to provide high-quality services to mobile users.

\textbf{Agile network management}. Through dedicated control channel, network operator obtains timely global information. Via open programmable interfaces, informed decisions and policies can be dynamically enforced to the network components. Being separated from the underlying hardware, network management and orchestration are totally software based, which makes high-level operational automation possible. In this way, the network operators are free of endless configuration, aimless debugging and maintenance.

\section{Case Studies}\label{cases}
In this section, the feasibility and advantage of the SDN enabled architecture are illuminated by two case studies: crowdsourcing task scheduling and popularity-aware content caching.

\subsection{Crowdsourcing Task Scheduling}
\begin{figure}[]
  \centering
  \includegraphics[width=0.48\textwidth]{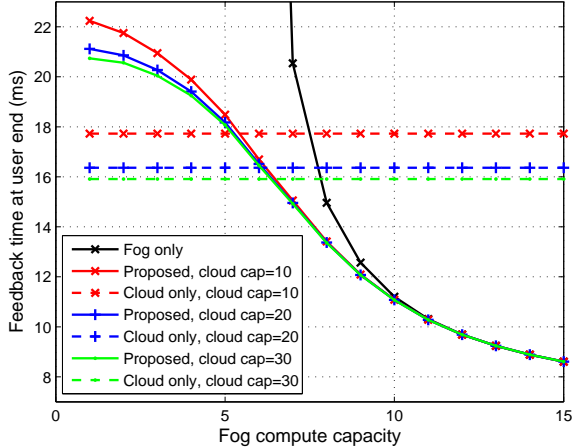}
  \caption{Average feedback time at user end versus fog compute capacity. The colored solid lines illustrate the results under the proposed architecture, while the black solid line indicates the results of fog only scenario, dashed lines depict the results of cloud only scenario.}\label{responsetimefogcapacity}
\end{figure}

Crowdsourcing is a new paradigm that solicits collective collaboration to support applications that require either broad participation or high performance computation \cite{crowdsourcing}. Those requirements generally exceed the capacity of any single agent. As such, multiple agent needs to contribute collectively. For example, neighbouring fog servers on the road need to cooperate to perform road traffic prediction \cite{vfog}. In those scenarios, MC needs decompose and distribute tasks to adjacent fog servers in a crowdsourcing manner. So in this study, we evaluate how SDN enabled cloud-fog interoperation can improve the responsiveness of computational crowdsourcing tasks.

We make the following assumptions: 1) The local crowdsourcing tasks is processed in First In, First Out (FIFO) manner; 2) The cloud starts processing the task immediately upon receipt and 3) MC makes scheduling decisions according to the following rule: if the estimated fog response time\footnote{Defined as the duration between the first run time and arrival time.} of crowdsourcing is shorter than the request delivery time to the cloud, it decomposes the task and put it in the FIFO queue, otherwise it sends the request to the cloud.

The measurements of task workload and fog/cloud compute capacity are normalized. Assume MC processes $1000$ crowdsourcing requests, arriving according to Poisson process with parameter $0.25$, and each with uniformly distributed workload between $0$ and $40$. We measure the feedback time at user end (including the round-trip time for request delivery, the response time and the processing time) under different scenarios, so as to evaluate the performance of the proposed architecture. The round-trip time of local crowdsourcing and cloud computing are set to $6$ ms and $15$ ms, respectively \cite{GSMA Latency}.

Fig. \ref{responsetimefogcapacity} illustrates the the trends of how the average feedback time at user end varying with the increase of compute resources at fog. It can be seen that, under the proposed architecture, the average feedback time at user end drops rapidly below the cloud only scenario. When fog compute capacity is higher than $5$, it is capable of processing most of the requests, so the cloud capacity has little impact on the feedback time. The results under proposed architecture converge to the fog only scenario indicates that, fog is able to process all the requests after its compute capacity exceeds $10$. The benefit of the proposed scheme can be even more striking if the proportion of lightweight tasks increases.

This study confirms that the proposed architecture can well catalyze the cloud-fog interoperation, offering better QoE with lightweight fog resources (compute capacity between $5$ and $10$ in this case). The observation is that, the MC should keep the real-time information of available local resources so as to make optimal scheduling decisions. Meanwhile, more fog resources will absolutely improve QoE, but the capital expenditure and operational expenditure will also increase. Thus, the insight of this study can be used to instruct fog resources deploying in practice.

\subsection{Popularity-aware Content Caching}
\begin{figure*}
\centering
\subfigure[]{\label{latencycachingsize}
\includegraphics[width=0.48\textwidth]{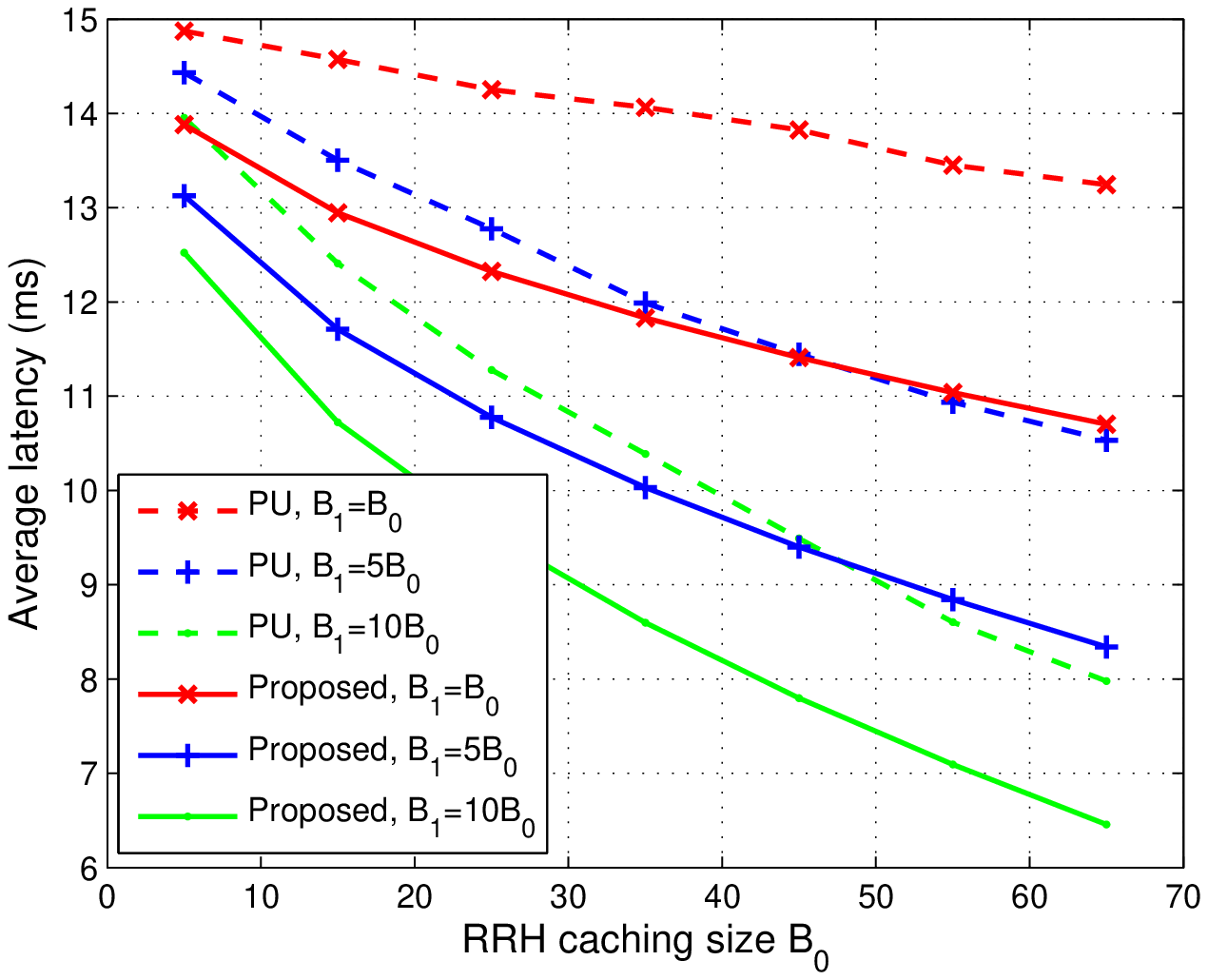}}
\subfigure[]{\label{trafficcachingsize}
\includegraphics[width=0.48\textwidth]{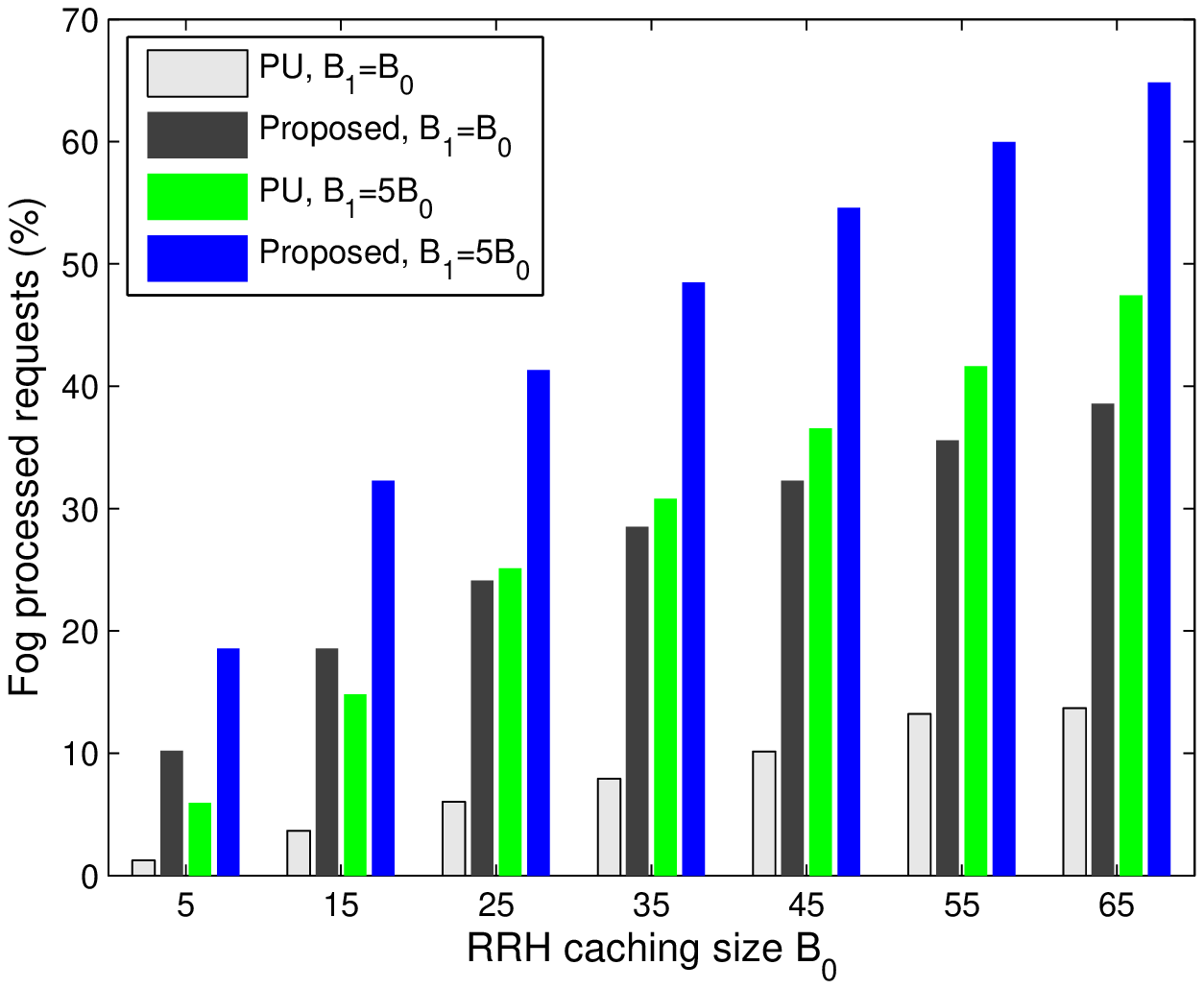}}
\caption{The comparison between interoperation enabled caching strategy and popularity unaware (PU) caching. (a) Average latency versus the RRH caching size, (b) Percentage of fog processed requests versus the RRH caching size.}
\label{Sim_result}
\end{figure*}

Content sharing, especially video streaming, will generate the majority of mobile traffic in the future \cite{cisco VNI}. In this study, we target on content caching, and evaluate how the SDN enabled architecture improves QoE and, meanwhile, alleviate the backhaul bandwidth pressure.

Considering the following scenario: mobile users continuously request video clips from content servers. Without interoperation, fog server can only cache contents in a popularity-unaware manner, such as caching the latest requested contents. While in the proposed architecture, since user requests are properly recorded in the system log, the controller can easily identify the hot video contents and cache them from cloud.

We consider a medium-sized urban network where a BBU supports $100$ RRHs \cite{cran tutorial}, and each RRH serves $500$ mobile users. Assume there are total $K=1000$ video clips in the cloud server. Since we can crop the large files into small pieces, all video clips are considered with unit size for simplicity. Their popularity distribution follows Zipf's law with parameter $\alpha$, i.e., the content ranking the $i$th is requested with frequency $f_i= c/i^{\alpha}$, $i=1, . . . , K$. Set $\alpha=0.56$ \cite{mobihoc} and let the normalization constant $c$ equal to $1$. Assume the storage size of RRH is $B_0$, and that of BBU is $B_1$. We assume the round-trip communication from mobile users to its nearest RRH fog, BBU fog and the cloud is $3$ ms, $6$ ms and $15$ ms \cite{GSMA Latency} respectively. As interoperation is realized, we employ the following straightforward caching policy: the controller always caches the most popular $B_0$ contents from cloud at RRH while caches the less popular $B_1$ ones at BBU. The requests fail to be met at RRH or BBU are sent to BBU or cloud accordingly.

Fig. \ref{latencycachingsize} shows that compared to the popularity unaware scenario, our interoperation enabled strategy provides better latency performance. Fig. \ref{trafficcachingsize} shows that fog servers are able to process more user requests under the proposed architecture, which, in turn, reduces more traffic traversed the backhaul. Again, this study confirms the advantage of the proposed architecture. It can be seen that, popularity-aware caching under our architecture can effectively improve QoE and alleviate backhaul pressure, especially under the scenario with less fog resources ($B_1=B_0$).

Moreover, as the popularity of video clips changes over time, contents in fogs have to be dynamically added or evicted. By employing machine learning techniques in the system log, the logical controller can capture the ascending or descending trends of certain contents, thus proactive adding or evicting instructions can be made to provide better QoE with limited resources.

\subsection{Remarks}
As we can see from the case studies, the proposed architecture can not only improve the QoE, but also benefit the network operator in terms of mitigated backhaul pressure and agile network management. However, the cost for deploying this promising architecture should also be considered. Firstly, network operator will have to investigate more in deploying initial fog infrastructures, reskilling the stuff and customizing new business model within the new architecture. Secondly, SPs will also have to investigate the possible candidate services that can be better delivered from the fog, and well balance the weight between fog and cloud for reliability, responsiveness and cost effectiveness concerns.

\section{Potential Research Issues}\label{potential issues}
Leveraging SDN for cloud-fog interoperation is promising yet challenging. In this section, we identify several specific research issues that can be further investigated to unleash the potential of the proposed architecture:
\begin{enumerate}
   \item \textbf{Security and privacy}. Generally, fog is more secure than cloud, since user requests are processed locally and the chance of user data exposure is reduced. However, security and privacy issues also arise from several aspects. Firstly, the centralized SDN paradigm may suffer from attacks on the control channel and the controller, where the network could be fatally comprised \cite{sdnsurvey,securityMag}. In this way, designing secure control channel and reliable controller is of paramount importance. Secondly, user's privacy also at stake, especially in the crowdsourcing applications \cite{crowdsourcing}. As MC distributes user requests to multiple fog servers, users' private information disclosure at fog server will be inevitable. Hence, designing secure and reliable communication strategies is non-trivial.
  \item \textbf{Cloud-fog resource balancing}. The case studies indicates that mounting fogs with larger compute and storage capacity will significantly boost the performance. But the consequent capital expenditure and operational expenditure also matters. In this way, we should investigate the optimal amount of fog resources that need to be deployed, especially when users' demands are fluctuating.
  \item \textbf{Controller design}. The controller should be knowledgable and intelligent. With the coordination of MC, the controllers are expected to work cooperatively so that the limited fog resources can be efficiently utilized. Moreover, as system log expands, learning pattern from the history statistics can well serve the purpose of traffic prediction and dynamic resource allocation. The controller can even capture the pattern of users' habit of accessing the network. Certain APIs should be designed to support controller's intelligence at different levels.
  \item \textbf{Cooperation among network operators}. Fogs are necessarily managed by different network operators. They need to cooperate closely to make the fog-fog and cloud-fog interoperation possible. Optimal cooperation model and incentive mechanisms that stimulate more participation of operators should be designed.
\end{enumerate}

\section{Conclusion}\label{conclusion}
In this article, we have presented an SDN enabled architecture for cloud-fog interoperation, which directly contributes to improved QoE and agile network management. We first describe the necessity of fog computing by arguing that, cloud computing will introduce unwilling latency and backhaul bottleneck and it is not well-suited for location based service provisioning. We then present how SDN controller can be exploited to bridge the interoperation of cloud and fog computing. Two case studies are discussed to illuminate the feasibility and advantage of the SDN enabled cloud-fog architecture. Potential research issues are given to further harness SDN for cloud-fog interoperation.

\ifCLASSOPTIONcaptionsoff
  \newpage
\fi

\flushend
\end{document}